\documentclass[twocolumn,amsmath,
  amssymb,english,aps,prb,floatfix,showpacs]{revtex4}
\usepackage[english]{babel}
\usepackage{times}
\makeatletter

\usepackage{graphicx}

\usepackage{babel}
\makeatother
\begin{document}

\title{Thermal entanglement of qubit pairs on the Shastry-Sutherland lattice}

\author{S. El Shawish}

\affiliation{J. Stefan Institute, Ljubljana, Slovenia}

\author{A. Ram\v{s}ak}

\affiliation{Faculty of Mathematics and Physics, University of Ljubljana, Ljubljana,
Slovenia}

\affiliation{J. Stefan Institute, Ljubljana, Slovenia}

\author{J. Bon\v{c}a}

\affiliation{Faculty of Mathematics and Physics, University of Ljubljana, Ljubljana,
Slovenia}

\affiliation{J. Stefan Institute, Ljubljana, Slovenia}

\date{2 April 2007}

\begin{abstract}
We show that temperature and magnetic field properties of the
entanglement between spins on the two-dimensional Shastry-Sutherland
lattice can be qualitatively described by analytical results for a
qubit tetramer. Exact diagonalization of clusters with up to 20 sites
reveals that the regime of fully entangled neighboring pairs coincides
with the regime of finite spin gap in the spectrum. Additionally, the
results for the regime of vanishing spin gap are discussed and related
to the Heisenberg limit of the model.
\end{abstract}


\maketitle

\section{Introduction}

In any physical system with subsystems in interaction, individual
parts of the system are to some extent entangled, even if they are far
apart, as realized already at the beginning of modern quantum
mechanics sixty years ago. Today it has become appreciated that the
ability to establish entanglement between quantum particles in a
controlled manner is a crucial ingredient of any quantum information
processing system\cite{nielsen01}. On the other hand, it turned out
that the analysis of appropriately quantified entanglement between
parts of the system can also be a very useful tool in the study of
many body phenomena, as is, {\it e.g.}, the behavior of correlated
systems in the vicinity of crossovers between various regimes or even
points of quantum phase transition\cite{osterloh02}.

Quantum entanglement of two distinguishable particles in a pure state
can be quantified through von Neuman entropy
\cite{bennett96,hill97,vedral97}. Entanglement between two
spin-${1 \over 2}$ particles -- qubit pair -- can be considered a
physical resource, an essential ingredient of algorithms suitable for
quantum computation. For a pair of subsystems A and B, each occupied
by a single electron, an appropriate entanglement measure is the
entanglement of formation, which can be quantified from the Wootters
formula\cite{wootters98}. In general, electron-qubits have the
potential for even richer variety of entanglement measure choices due
to both their charge and spin degrees of freedom. When entanglement is
quantified in systems of indistinguishable particles, the measure must
account for the effect of exchange and it must adequately deal with
multiple occupancy states
\cite{schliemann01,ghirardi04,eckert02,gittings02,zanardi02,vedral03}.
A typical  example is the analysis of entanglement in lattice
fermion models (the Hubbard model, e.g.)  where double occupancy
plays an essential role\cite{zanardi02}.

In realistic hardware designed for quantum information processing,
several criteria for qubits must be fulfilled\cite{criteria}: the
existence of multiple identifiable qubits, the ability to initialize
and manipulate qubits, small decoherence, and the ability to measure
qubits, {\it i.e.}, to determine the outcome of computation. It seems that among
several proposals for experimental realizations of such quantum
information processing systems  the criteria for scalable
qubits can be met in solid state structures consisting of coupled
quantum dots\cite{divincenzo05,coish06}. Due to the ability to
precisely control the number of electrons in such structures
\cite{elzerman03}, the entanglement has become experimentally
accessible quantity. In particular, recent experiments on
semiconductor double quantum dot devices have shown the evidence of
spin entangled states in GaAs based heterostuctures\cite{chen04} and
it was shown that vertical-lateral double quantum dots may be useful
for achieving two-electron spin entanglement\cite{hatano05}. It was
also demonstrated recently that in double quantum dot systems coherent
qubit manipulation and projective readout is possible \cite{petta05}.

Qubit pairs to be used for quantum information processing must be to a
high degree isolated from their environment, otherwise small
decoherence requirement from the DiVincenzo's checklist can not be
fulfilled. The entanglement, {\it e.g.}, between two
antiferromagnetically coupled spins in contact with thermal bath, is
decreased at elevated temperatures and external magnetic
field\cite{nielsen00,arnesen01,wang01}, and will inevitably vanish at
some finite temperature\cite{fine05}.
Entanglement of a pair of electrons that are confined in a double
quantum dot is collapsed due to the Kondo effect at low
temperatures and for a very weak tunneling to the leads.
%
%
At temperatures below the Kondo temperature a spin-singlet state
is formed between a confined electron and conduction electrons in
the leads\cite{rmzb06}. For other open systems there are many
possible sources of decoherence or phase-breaking, for example
coupling to phonon degrees of freedom \cite{yu02}.

The main purpose of the present paper is to analyze the robustness of
the entanglement of spin qubit pairs in a planar lattice of spins
(qubits) with respect to frustration in magnetic couplings, elevated
temperatures as well as due to increasing external magnetic field. The
paper is organized as follows.  Sec. \ref{secII} introduces the
model for two coupled qubit pairs -- qubit tetramer -- and presents
exact results for temperature and magnetic field dependence of the
entanglement between nearest and next-nearest-neighboring spins in a
tetrahedron topology.  In Sec. \ref{secIII} the model is extended to
infinite lattice of qubit pairs described by the Shastry-Sutherland
model \cite{shastry81}. This model is convenient firstly, because of
the existence of stable spin-singlet pairs in the ground state in the
limit of weak coupling between the qubit pairs, and secondly, due to a
relatively good understanding of the physics of the model in the
thermodynamic limit. Entanglement properties of the Shastry-Sutherland
model were so far not considered  quantitatively. Neverteless, several
results concerning the role of entanglement at a phase
transition in other low-dimensional spin lattice systems
\cite{osterloh02,syljuasen03,amico04,osborne02,roscilde04,roscilde05,larsson05},
as well as in fermionic systems\cite{gu04,deng05,legeza06} have
been reported recently. Near a quantum phase transition in some
cases entanglement even proves to be more efficient precursor of
the transition compared to standard spin-spin correlations
\cite{verstraete04,legeza06}. In Sec. \ref{secIV} we discuss
entanglement between nearest neighbors in the Heisenberg model,
representing a limiting case of the Shastry-Sutherland model. Results
are summarized in Sec. \ref{secV} and some technical details are given
in Appendix A.

\section{Thermal entanglement of a qubit tetramer in magnetic field}
\label{secII}

\begin{figure}
\begin{center}
\includegraphics[width=60mm,keepaspectratio]{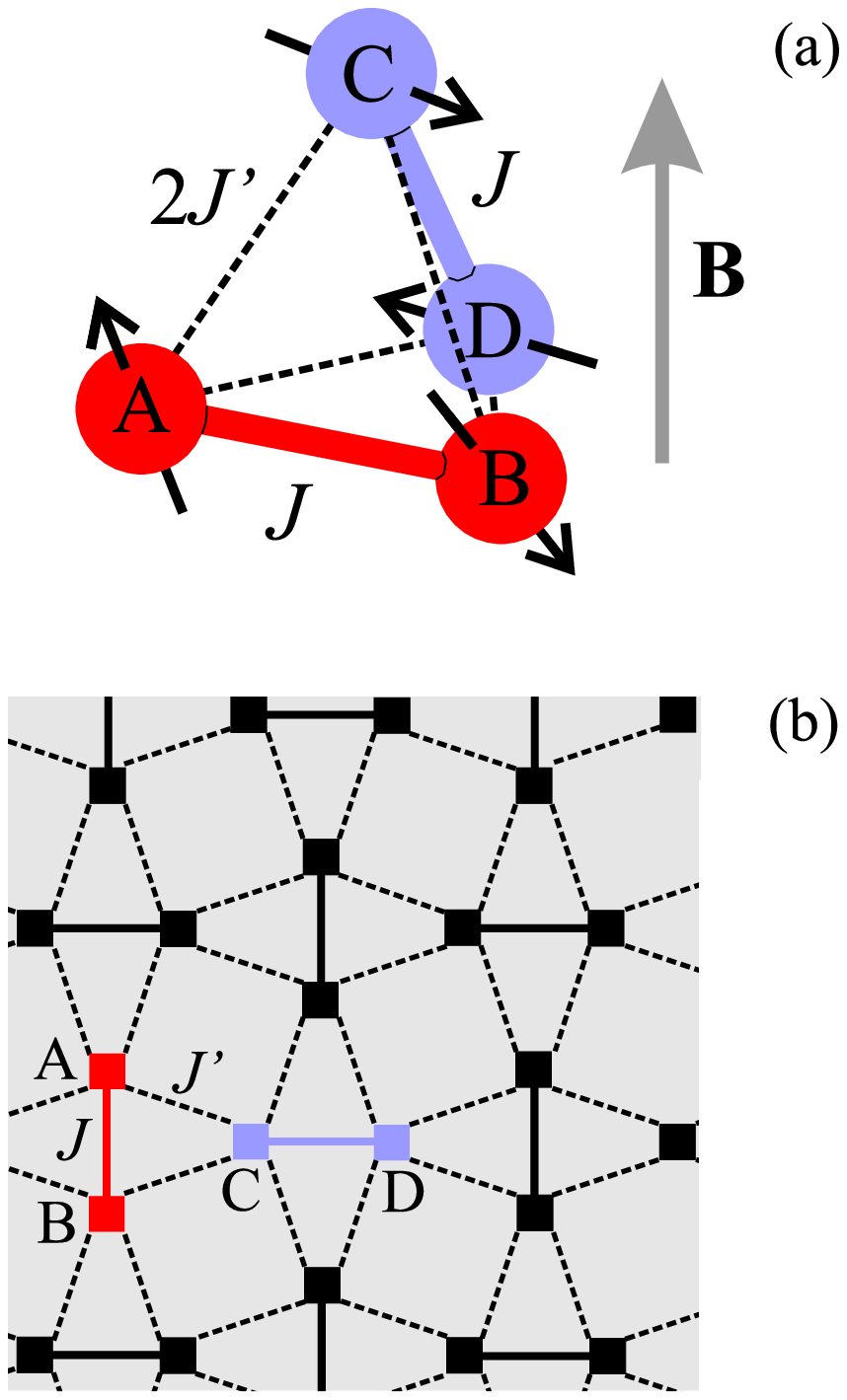}
\end{center}

\caption{(Color online) (a) Two coupled qubit pairs (dimers) in
tetrahedral topology. (b) Shastry-Sutherland lattice as realized, {\it e.g.},
in the SrCu$_{2}$(BO$_{3}$)$_{2}$ compound.}

\label{fig1}
\end{figure}

Consider first a double quantum dot composed of two adjacent quantum
dots weakly coupled via a controllable electron-hopping integral. By
adjusting a global back-gate voltage, precisely two electrons can be
confined to the dots.  The inter-dot tunneling matrix element $t$
determines the effective antiferromagnetic (AFM) superexchange
interaction $J\sim 4t^{2}/U$, where $U$ is the scale of Coulomb
interaction between two electrons confined on the same dot. There are
several possible configurations of coupling between such double
quantum dots. One of the simplest specific designs is shown
schematically in Fig.~\ref{fig1}(a): four qubits at vertices of a
tetrahedron. In addition to the coupling A-B, by appropriate
arrangements of gate electrodes the tunneling between A-C and A-D can
as well be switched on.

We consider here the case where $J/U \ll 1$, thus double occupancy of
individual dot is negligible and appropriate Hilbert space is spanned
by two dimers (qubit pairs): spins at 
sites A-B and C-D
are coupled by effective AFM Heisenberg magnetic exchange $J$ and at sites
A-C, B-C, A-D, B-D by $J'$. The corresponding
hamiltonian of such a pair of dimers is given as
\begin{eqnarray}
\label{habcd}
H_4&=&J({\textbf{S}}_{\textrm{A}}\cdot{\textbf{S}}_{\textrm{B}}+
{\textbf{S}}_{\textrm{C}}\cdot{\textbf{S}}_{\textrm{D}})+\\ \nonumber
&+&2J^\prime({\textbf{S}}_{\textrm{A}}\cdot{\textbf{S}}_{\textrm{C}}+
{\textbf{S}}_{\textrm{B}}\cdot{\textbf{S}}_{\textrm{C}}+
{\textbf{S}}_{\textrm{A}}\cdot{\textbf{S}}_{\textrm{D}}+
{\textbf{S}}_{\textrm{B}}\cdot{\textbf{S}}_{\textrm{D}})- \\ \nonumber
&-&B(S^z_{\textrm{A}}+S^z_{\textrm{B}}+S^z_{\textrm{C}}+S^z_{\textrm{D}}),  \nonumber
\end{eqnarray}
where $\textbf{S}_i={1 \over 2} {\bf \sigma}_i$ is spin operator
corresponding to the site $i$ and $B$ is external homogeneous magnetic
field in the direction of the $z$-axis. Factor 2 in Eq.~(\ref{habcd})
is introduced for convenience -- such a parameterization represents
the simplest case of finite Shastry-Sutherland lattice with periodic
boundary conditions studied in Sec. \ref{secIII}.

\begin{figure*}
\begin{center}
\includegraphics[width=60mm,keepaspectratio]{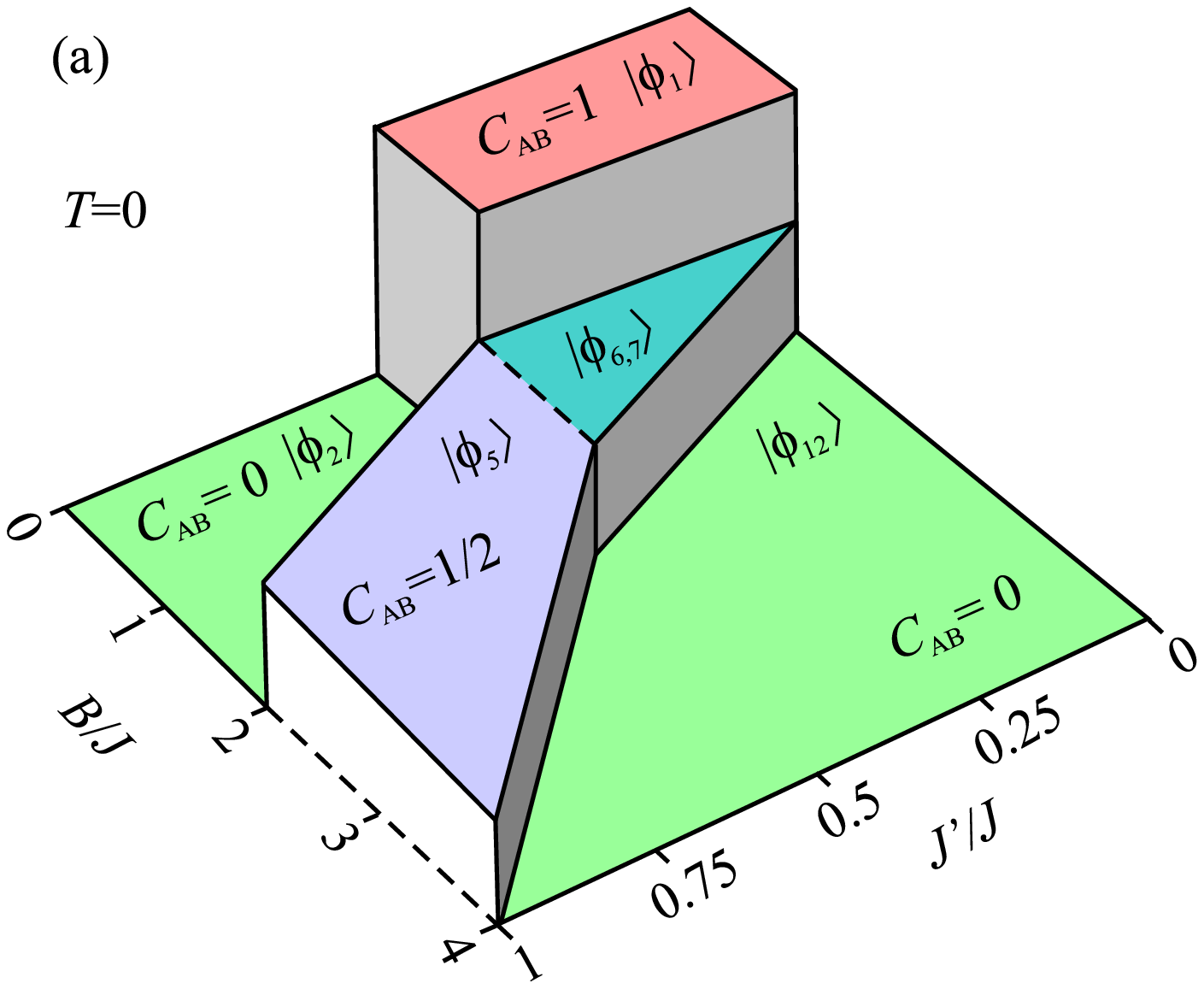}
\includegraphics[width=60mm,keepaspectratio]{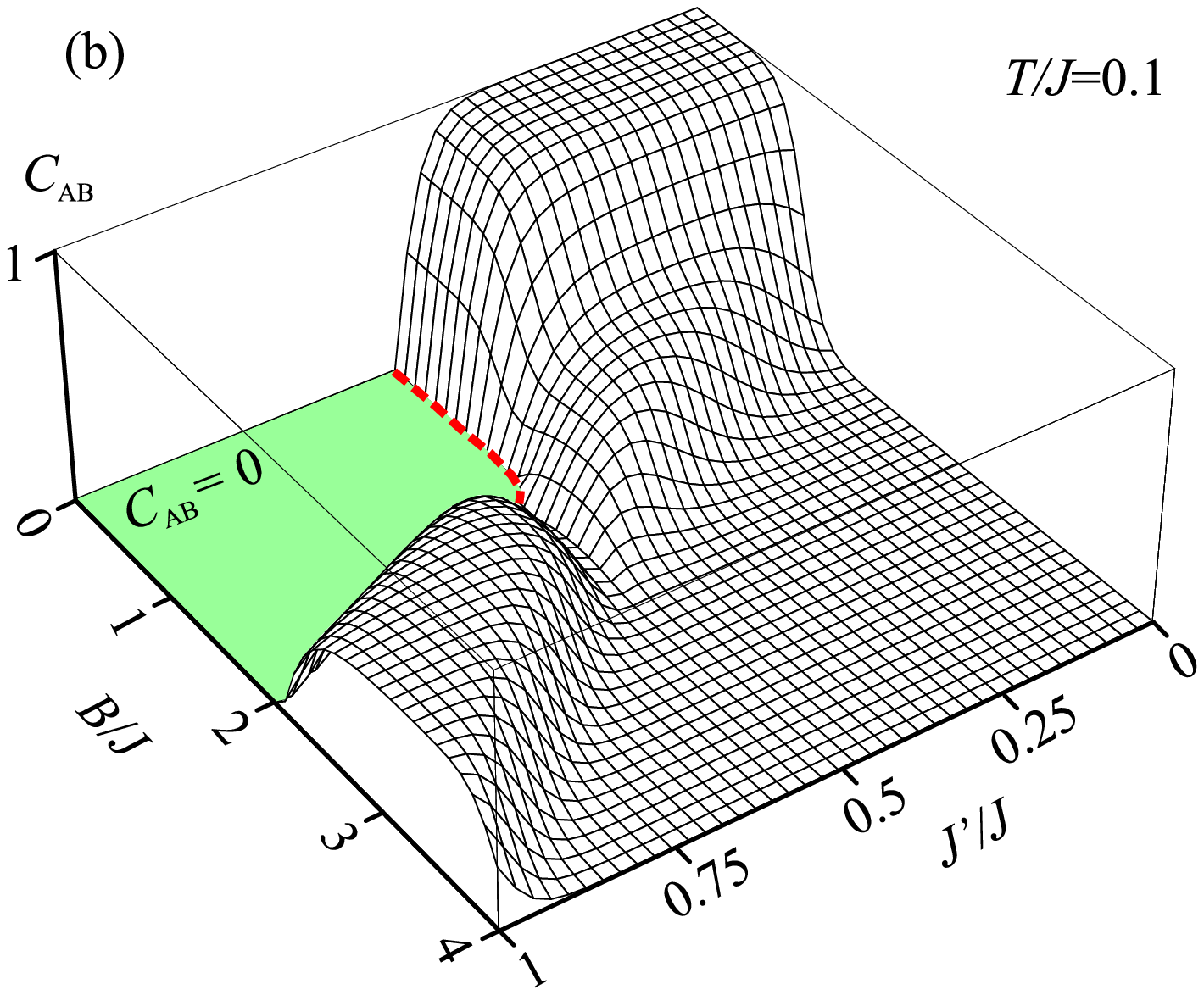}\\
\includegraphics[width=60mm,keepaspectratio]{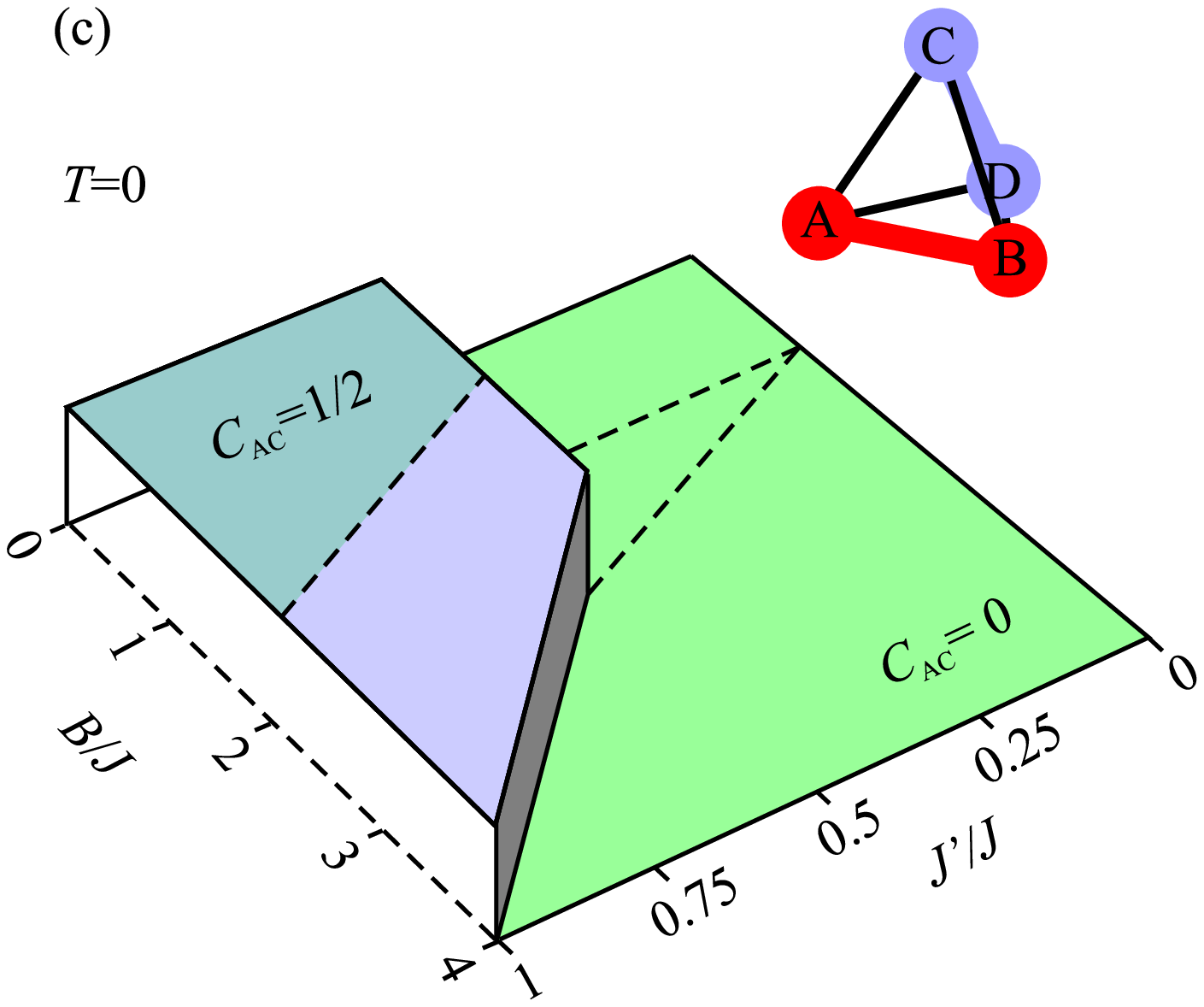}
\includegraphics[width=60mm,keepaspectratio]{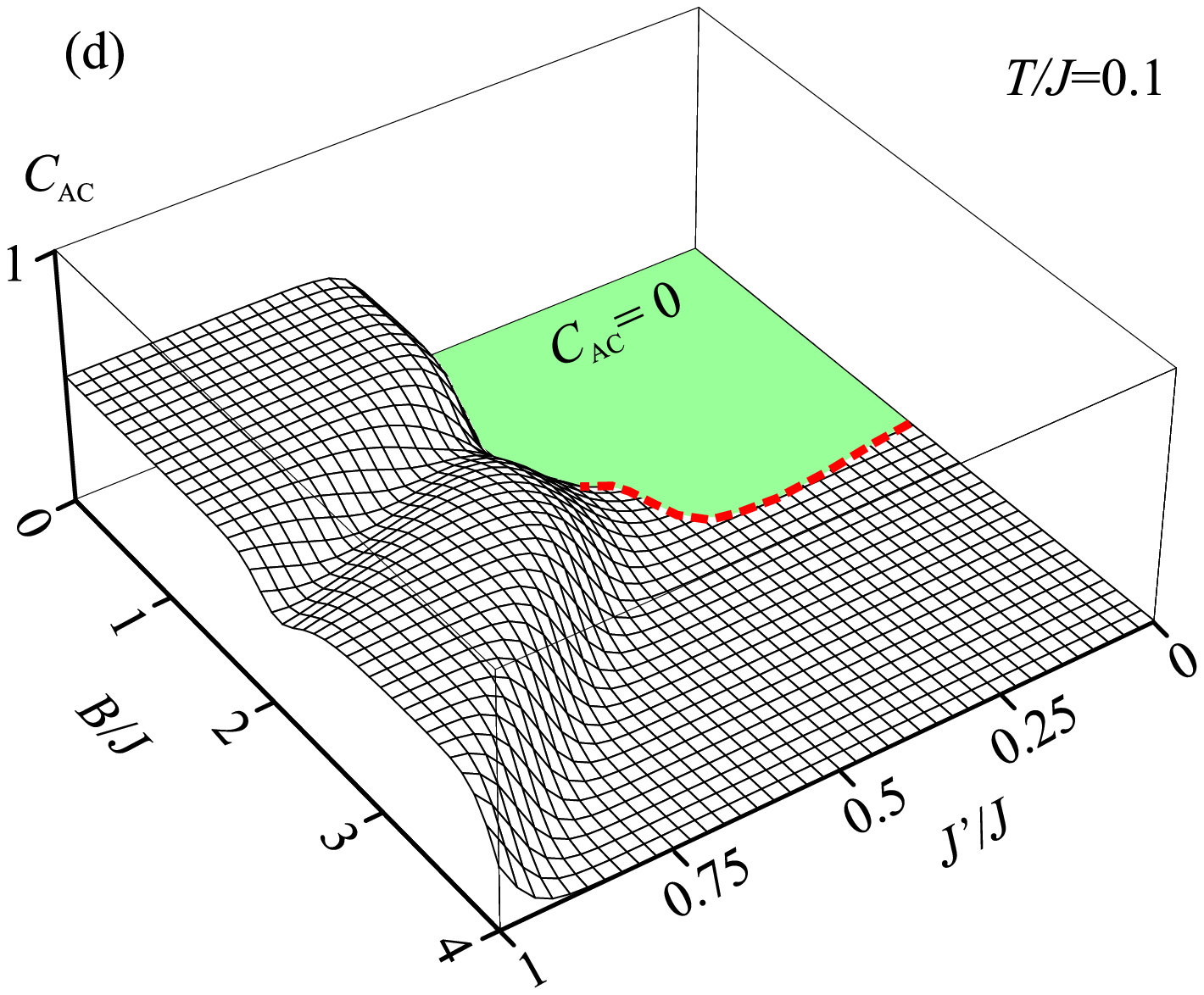}
\end{center}

\caption{(Color online) (a) Zero-temperature concurrence
$C_{\mathrm{A} \mathrm{B}}$ as a function of $J'/J$ and $B/J$.
Different regimes are characterized by particular ground state
functions $ |\phi_n\rangle$ defined in Appendix A.  (b) $T/J=0.1$
results for $C_{\mathrm{A} \mathrm{B} }$. (c) Next nearest concurrence
$C_{\mathrm{A} \mathrm{C} }$ for $T=0$, and (d) for $T/J=0.1$.
Dashed lines separate $C_{\mathrm{A}\mathrm{B}(\mathrm{C})}>0$ from
$C_{\mathrm{A}\mathrm{B}(\mathrm{C})}=0$.}

\label{fig2}
\end{figure*}

\subsection{Concurrence}

We focus here on the entanglement properties of two coupled qubit
dimers. The entanglement of a pair of spin qubits A and B may be
defined through concurrence\cite{bennett96}, $C =
2|\alpha_{\uparrow\!\uparrow}\alpha_{\downarrow\!\downarrow}-
\alpha_{\uparrow\!\downarrow}\alpha_{\downarrow\!\uparrow}|$, if the
system is in a pure state
$|\Psi_{\mathrm{AB}}\rangle=\sum_{ss'}\alpha_{ss'}|s\rangle_{\!
\mathrm{A}} |s'\rangle_{\! \mathrm{B}}$, where $|s \rangle_i$
corresponds to the basis $|\!\uparrow\,\rangle_i$,
$|\!\downarrow\,\rangle_i$. Concurrence varies from $C=0$ for an
unentangled state (for example $|\!\uparrow\,\rangle_\mathrm{A}
|\!\uparrow\,\rangle_\mathrm{B}$) to $C=1$ for completely entangled
Bell states\cite{bennett96} $ {1 \over \sqrt{2}}(
|\!\uparrow\,\rangle_\mathrm{A} |\!\uparrow\,\rangle_\mathrm{B} \pm
|\!\downarrow\,\rangle_\mathrm{A} |\!\downarrow\,\rangle_\mathrm{B})$
or ${1 \over \sqrt{2}}(|\!\uparrow\,\rangle_\mathrm{A}
|\!\downarrow\,\rangle_\mathrm{B} \pm
|\!\downarrow\,\rangle_\mathrm{A} |\!\uparrow\,\rangle_\mathrm{B})$.

For finite inter-pair coupling $J^\prime \ne 0$ or at elevated
temperatures the A-B pair can not be described by a pure state. In the
case of mixed states describing the subsystem A-B the concurrence may
be calculated from the reduced density matrix
$\rho_{\textrm{A}\textrm{B}}$ given in the standard basis $|s
\rangle_i |s' \rangle_j$\cite{wootters98}.  Concurrence can be further
expressed in terms of spin-spin correlation functions
\cite{osterloh02,syljuasen03}, where for systems that are axially
symmetric in the spin space the concurrence may conveniently be given in a
simple closed form\cite{ramsak06}, which for the thermal
equilibrium case simplifies further,
\begin{equation}
C_{\mathrm{A}  \mathrm{B}  } =2 \textrm{max}(0,
|\langle S_{\mathrm{A}}^{+}
S_{\mathrm{B}}^{-}\rangle|-\sqrt{\langle P_{\mathrm{A}}^
{\uparrow}P_{\mathrm{B}}^{\uparrow}\rangle\langle P_{\mathrm{A}}^
{\downarrow}P_{\mathrm{B}}^{\downarrow}\rangle}).
\label{cmax}
\end{equation}
Here $S_i^{+} = (S_i^{-})^{\dagger} = S_i^x+\imath S_i^y$ is the spin
raising operator for dot $i$ and $P_i^\uparrow={1 \over 2}(1+2
S_i^z)$, $P_i^\downarrow={1 \over 2}(1-2 S_i^z)$ are the projection
operators onto the state $|\!\uparrow\,\rangle_i$ or
$|\!\downarrow\,\rangle_i$, respectively. We consider the concurrence
at fixed temperature, therefore the expectation values in the
concurrence formula Eq.~(\ref{cmax}) are evaluated as
\begin{equation}
\langle {\cal O} \rangle = {1 \over Z} \sum_n \langle n| {\cal O}|n \rangle e^{-\beta E_n},
\label{o}
\end{equation}
where $Z= \sum_n e^{-\beta E_n}$ is the partition function,
$\beta=1/T$, and $\{|n \rangle \}$ is a complete set of states of the
system. Note that due to the equilibrium and symmetries of the system,
several spin-spin correlation functions vanish, $\langle
S_{\mathrm{A}}^{+}S_{\mathrm{B}}^{+}\rangle=0$, for example.

In vanishing magnetic field, where the SU(2) symmetry is restored, the
concurrence formula Eq.~(\ref{cmax}) simplifies further and is
completely determined by only one\cite{werner89} spin invariant
$\langle\textrm{\bf S}_{\mathrm{A}}\cdot\textrm{\bf
S}_{\mathrm{B}}\rangle$,
\begin{equation}
C_{\mathrm{A}  \mathrm{B}  } = \textrm{max}(0,-2\langle\textrm{\bf
S}_{\mathrm{A}}\cdot\textrm{\bf
S}_{\mathrm{B}}\rangle-\frac{1}{2}).
\label{su2}
\end{equation}
The concurrence may be expected to be significant whenever enhanced
spin-spin correlations indicate A-B singlet formation.

\subsection{Analytical results}

There are several known results related to the model
Eq.~(\ref{habcd}).  In the special case of $J^\prime=0$, for example,
the tetramer consists of two decoupled spin dimers with concurrence
$C_{\mathrm{A} \mathrm{B}}$ (or the corresponding thermal
entanglement) as derived in Refs.~\onlinecite{nielsen00,arnesen01}.
Entanglement of a qubit pair described by the related XXZ Heisenberg
model with Dzyaloshinskii-Moriya anisotropic interaction can be also
obtained analytically\cite{wang01}.  Hamiltonian $H_4$ with additional
four-spin exchange interaction but in the absence of magnetic field
was considered recently in the various limiting cases\cite{bose05}.

Tetramer model Eq.~(\ref{habcd}) considered here is exactly solvable
and in Appendix A we present the corresponding eigenvectors and
eigenenergies.  The concurrence $C_{\mathrm{A}
\mathrm{B}}$ is for this case determined from Eq.~(\ref{cmax}) with
\begin{eqnarray}
\langle S_{\rm A}^+S_{\rm B}^-\rangle&=&\frac{1}{Z}\Big[\!\!
-e^{3j/2}/2-e^{j/2}\left(e^{b}+e^{-b}\right)/2
\nonumber\\
&+&e^{-j/2+4j'}/6
+e^{-j/2+2j'}\left(e^{b}+e^{-b}\right)/4
\nonumber\\
&+&e^{-j/2-2j'}\left(e^{b}/4+1/3
+e^{-b}/4\right)\!\!\Big],
\end{eqnarray}
\noindent
where $j=\beta J$, $j'=\beta J'$, $b=\beta B$, and with
\begin{eqnarray}
\langle P_{\rm A}^{\uparrow\downarrow}
P_{\rm B}^{\uparrow\downarrow}\rangle&=&\frac{1}{Z}\Big[
e^{j/2}\left(e^{\pm b}\right)+e^{-j/2+4j'}/3
\nonumber\\
&+&e^{-j/2+2j'}\left(1+e^{\pm b}\right)/2
\nonumber\\
&+&e^{-j/2-2j'}\left(1/6+e^{\pm b}/2+e^{\pm2 b}
\right)\!\!\Big].
\end{eqnarray}
Here
\begin{eqnarray}
Z&=&e^{3j/2}
+2e^{j/2}\left(e^{b}+1+e^{-b}\right)
\nonumber\\
&+&e^{-j/2+4j'}
+e^{-j/2+2j'}\left(e^{b}+1+e^{-b}\right)
\nonumber\\
&+&e^{-j/2-2j'}\left(e^{2b}+e^{b}+1
+e^{-b}+e^{-2b}\right)
\end{eqnarray}
is the partition function.

\begin{figure}
\begin{center}
\includegraphics[width=60mm,keepaspectratio]{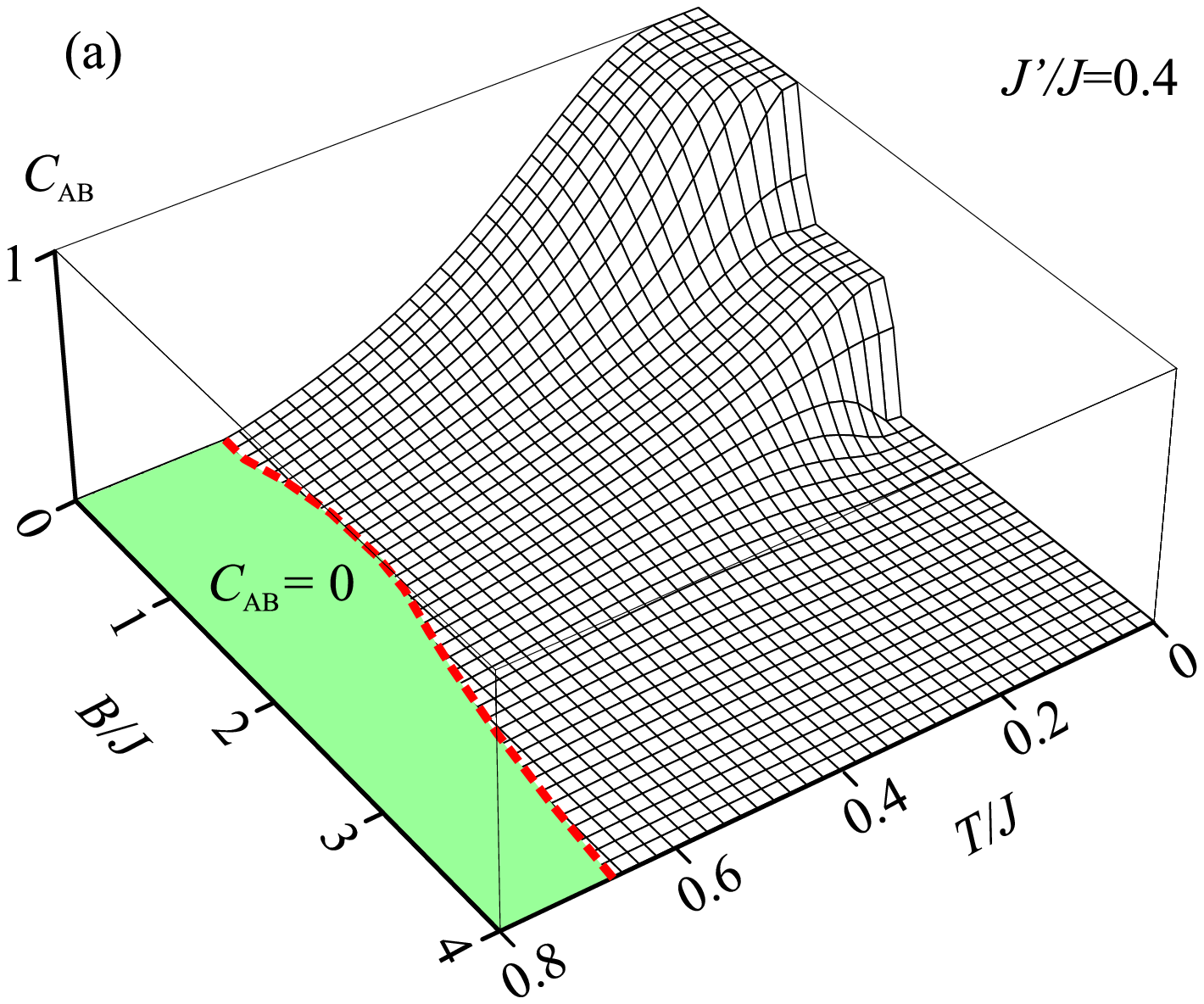}\\
\includegraphics[width=60mm,keepaspectratio]{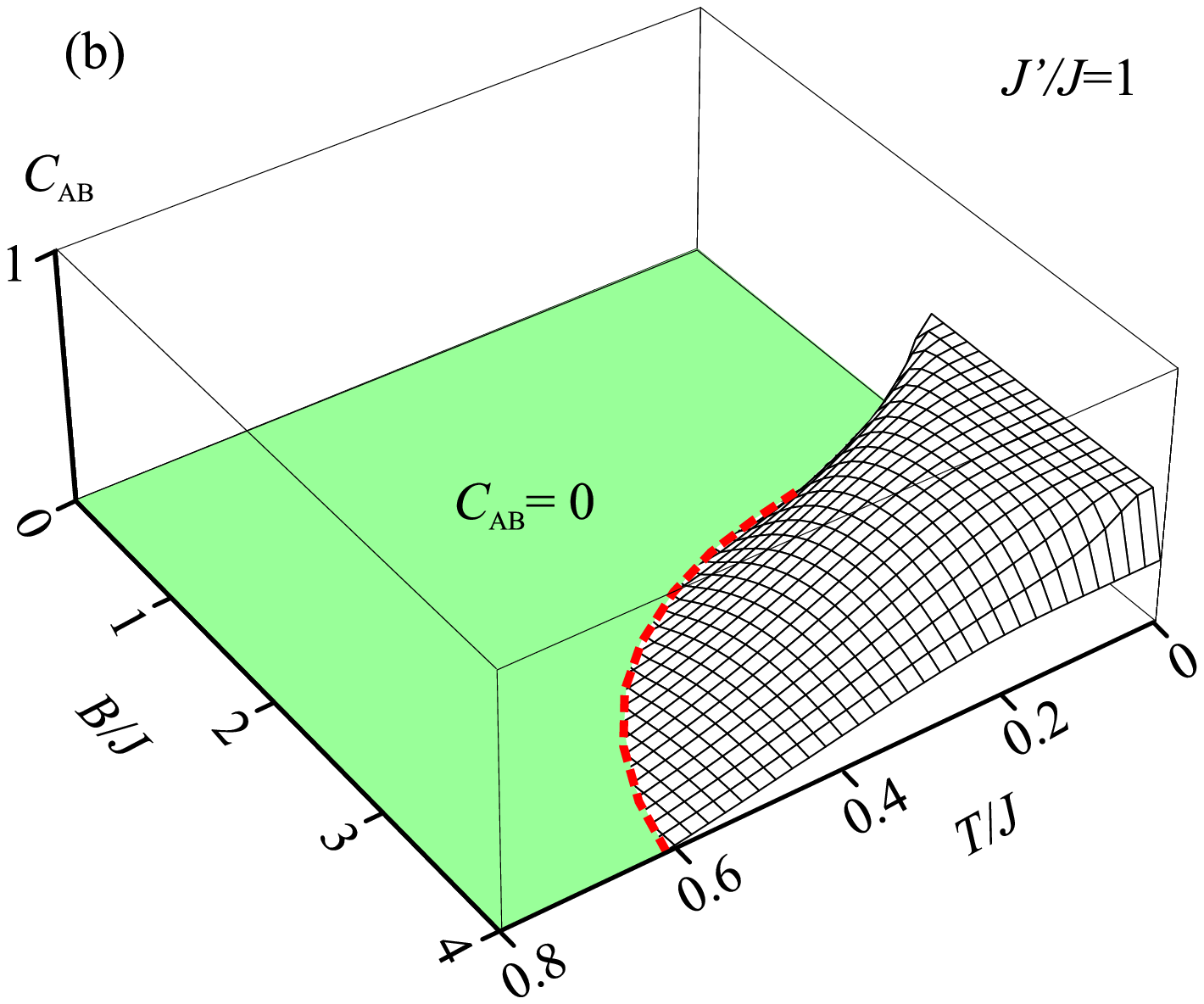}
\end{center}

\caption{(Color online) (a) Temperature and magnetic field dependence of
$C_{\mathrm{A}\mathrm{B}}$ for $J'/J=0.4$ and (b) $J'=J$.
Dashed lines separate $C_{\mathrm{A}\mathrm{B}}>0$ from
$C_{\mathrm{A}\mathrm{B}}=0$.
}

\label{fig3}
\end{figure}

Alternatively, one can define and analyze also the entanglement
between spins at sites A and C and the corresponding concurrence
$C_{\mathrm{A} \mathrm{C}}$ can be expressed from Eq.~(\ref{cmax}) by
applying additional correlators with replaced B$\to$C,
\begin{eqnarray}
\langle S_{\rm A}^+S_{\rm C}^-\rangle&=&\frac{1}{Z}\Big[\!\!
-e^{-j/2+4j'}/3
\nonumber\\
&-&e^{-j/2+2j'}\left(e^{b}+e^{-b}\right)/4
\nonumber\\
&+&e^{-j/2-2j'}\left(e^{b}/4+1/3
+e^{-b}/4\right)\!\!\Big],
\end{eqnarray}
and
\begin{eqnarray}
\langle P_{\rm A}^{\uparrow\downarrow}
P_{\rm C}^{\uparrow\downarrow}\rangle&=&\frac{1}{Z}\Big[
e^{3j/2}/4+e^{j/2}\left(1/2+e^{\pm b}\right)
\nonumber\\
&+&e^{-j/2+4j'}/12
+e^{-j/2+2j'} e^{\pm b}/2
\nonumber\\
&+&e^{-j/2-2j'}\left(1/6
+e^{\pm b}/2+e^{\pm2 b}\right)\!\!\Big].
\end{eqnarray}

The line $2J'=J$ represents a particularly interesting special case
where two dimers are
coupled symmetrically forming a regular tetrahedron. An important
property of this system is the (geometrical) frustration of, {\it
e.g.}, qubits C-A-B. Such a frustration is the driving force of the
quantum phase transition found in the Shastry-Sutherland model and is
the reason for similarity of the results for two coupled dimers and a
large planar lattice studied in the next Section.

\subsection{Examples}

In the low temperature limit the concurrence is determined by the
ground state properties while transitions between various regimes are
determined solely by crossings of eigenenergies, which depend on two
parameters $(J^\prime/J,B/J)$.  There are 5 distinct regimes for
$C_{\mathrm{A} \mathrm{B}}$ shown in Fig.~\ref{fig2}(a): (i)
completely entangled dimers (singlets A-B and C-D, state
$|\phi_1\rangle$ from Appendix A), $C_{\mathrm{A}
\mathrm{B}}=1$; (ii) for $B>J$ and smaller $J^\prime/J$ the
concurrence is zero because the energy of the state consisted of a
product of fully polarized A-B and C-D triplets, $|
\phi_{12}\rangle$, is the lowest energy in this regime; (iii)
concurrence is zero also for $J^\prime>J/2$ and low $B/J$, with the
ground state $| \phi_2 \rangle $.  There are two regimes corresponding
to ${1 \over 2}$ step in $C_{\mathrm{A} \mathrm{B}}$ where the ground
state is either (iv) any linear combination of degenerate states $|
\phi_{6,7}\rangle$, {\it i.e.}, simultaneous A-B singlet (triplet) and
C-D triplet (singlet) for $J^\prime<J/2$, or (v) state $|
\phi_5\rangle$ at $J^\prime>J/2$ and larger $B$. Qubits A-C are due to
special topology never fully entangled, and the corresponding
$C_{\mathrm{A} \mathrm{C}}$ is presented in Fig.~\ref{fig2}(c). In the
limit of $J^\prime \gg J$ the tetramer corresponds to a Heisenberg
model ring consisted of 4 spins and in this case qubit A is due to
tetramer symmetry equally entangled to both neighbors (C and D), thus
$C_{\mathrm{A} \mathrm{C}}={1 \over 2}$.

At elevated temperatures the concurrence is smeared out as shown in
Figs.~\ref{fig2}(b,d). Note the dip separating the two different
regimes with $C_{\mathrm{A} \mathrm{B}}={1 \over 2}$, seen also in the
$C_{\mathrm{A} \mathrm{C}}={1 \over 2}$ case. This dip clearly
separates different regimes discussed in the previous $T=0$ limit and
signals a proximity of a disentangled excited state. For sufficiently
high temperatures vanishing concurrence is expected
\cite{fine05}. The critical temperature $T_c$ denoted by a dashed line
is set by the magnetic exchange scale $J$, since
at higher temperatures local singlets are broken irrespectively of
the magnetic field.

A rather unexpected result is shown in Fig.~\ref{fig3}(a) where at
$B\gtrsim 2 J$ and low temperatures the concurrence slightly increases
with increasing temperature due to the contribution of excited A-B
singlet components that are absent in the ground state. Similar
behavior is found for $J^\prime\sim0$ around $B\sim J$, which is
equivalent to the case of a single qubit
dimer\cite{nielsen00,arnesen01} (not shown here). There is no
distinctive feature in temperature and magnetic field dependence of
$C_{\mathrm{A} \mathrm{B}}$ when $J'>J/2$ and a typical results is
shown in Fig.~\ref{fig3}(b) for $J'>J$.

\section{Planar array of qubit pairs: the Shastry-Sutherland lattice}
\label{secIII}
\subsection{Preliminaries}

The central point of this paper is the analysis of pair entanglement
for the case of a larger number of coupled qubit pairs. In the
following it will be shown that the results corresponding to tetramers
considered in the previous Section can be very helpful for better
understanding pair-entanglement of $N>4$ qubits. There are several
possible generalizations of coupled dimers and one of the simplest in
two dimensions is the Shastry-Sutherland lattice shown in
Fig.~\ref{fig1}(b).  Neighboring sites A-B are connected with exchange
interaction $J$ and next-neighbors with $J'$. The corresponding
hamiltonian for $N/2$ dimers ($N$ sites) is given with
\begin{equation}
H_N=J\sum_{\{\textrm{AB\}}}{\textbf{S}}_{i}\cdot{\textbf{S}}_{j}+
J'\sum_{\{\textrm{AC\}}}{\textbf{S}}_{i}\cdot{\textbf{S}}_{j}-B\sum_{i=1}^{N}S_{i}^{z}.
\end{equation}
Periodic boundary conditions are used. For the special case $N=4$ the
model reduces to Eq.~(\ref{habcd}) where due to periodic boundary
conditions sites A-C (and other equivalent pairs) are doubly
connected, therefore a factor of 2 in Eq.~(\ref{habcd}), as mentioned
in Sec. \ref{secII}.

The Shastry-Sutherland model (SSM) was initially proposed as a toy
model possessing an exact dimerized eigenstate known as a valence bond
crystal \cite{shastry81}. Recently, the model has experienced a sudden
revival of interest by the discovery of the two-dimensional
spin-liquid compound SrCu$_{2}$(BO$_{3}$)$_{2}$
\cite{smith91,kageyama99} since it is believed that magnetic properties
of this compound are reasonably well described by the SSM
\cite{miyahara03}. In fact, several generalizations of the SSM
have been introduced to account better for recent high-resolution
measurements revealing the magnetic fine structure of
SrCu$_{2}$(BO$_{3}$)$_{2}$
\cite{miyahara03,jorge03,elshawish1,elshawish2}. Soon after the
discovery of the SrCu$_{2}$(BO$_{3}$)$_{2}$ system, the SSM thus
became a focal point of theoretical investigations in the field of
frustrated AFM spin systems, particularly low-dimensional quantum
spin systems where quantum fluctuations lead to magnetically
disordered ground states (spin liquids) with a spin gap in the
excitation spectrum.

The SSM is a two-dimensional frustrated antiferromagnet with a unique
spin-rotation invariant exchange topology that leads in the limit
$J\gg J'$ to an exact gapped dimerized ground state with localized
spin singlets on the dimer bonds (dimer phase). In the opposite limit,
$J\ll J'$, the model becomes ordinary AFM Heisenberg model with a
long-range N\' eel order and a gapless spectrum (N\' eel phase). While
two of the phases are known, there are still open questions regarding
the existence and the nature of the intermediate phases. Several possible
scenarios have been proposed, {\it e.g.}: either a direct transition between the
two states occurs at the quantum critical point near $J'/J\sim 0.7$
\cite{miyahara99,mueller00}, or a transition via an intermediate phase
that exists somewhere in the range of $J'/J>0.6$ and $J'/J<0.9$
\cite{isacsson06}. Although different theoretical approaches have been
applied, a true nature of the intermediate phase (if any) has still
not been settled. As will be evident later on, our
exact-diagonalization results support the first scenario.

The SSM phase diagram reveals interesting behavior also for varying
external magnetic field. In particular, experiments on
SrCu$_{2}$(BO$_{3}$)$_{2}$ in strong magnetic fields show formation of
magnetization plateaus \cite{kageyama99,onizuka00}, which are believed
to be a consequence of repulsive interaction between almost localized
spin triplets.  Several theoretical approaches support the idea that
most of these plateaus are readily explained within the (bare) SSM
\cite{miyahara99,momoi00,misguich01}. Recent variational treatment
based on entangled spin pairs revealed new insight into various phases
of the SSM\cite{isacsson06}.

Although extensively studied, the zero-temperature phase diagram
of the SSM remains elusive. This lack of reliable solutions is
even more pronounced when considering thermal fluctuations in SSM
as only few methods allow for the inclusion of finite temperatures
in frustrated spin systems. In this respect, the calculation of
thermal entanglement between the spin pairs would also provide  a
new  insight into the complexity of the SSM.

\begin{figure}
\begin{center}
\includegraphics[width=70mm,keepaspectratio]{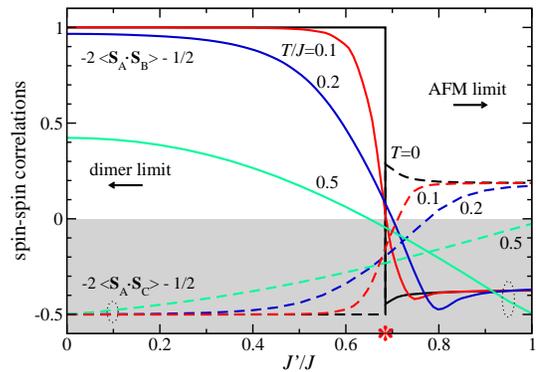}
\end{center}

\caption{(Color online) Results for the Shastry-Sutherland lattice with
$N=20$ sites and periodic boundary conditions. Presented are
renormalized spin-spin correlation functions $-2\langle\textrm{\bf
S}_{\mathrm{A}}\cdot\textrm{\bf S}_{\mathrm{B,C}}\rangle-\frac{1}{2}$
as a function of $J'/J$ and for various temperatures. Asterisk
indicates critical $J'_c$ which roughly separates the dimer and N\'
eel phase.}

\label{fig4}
\end{figure}

\begin{figure*}
\begin{center}
\includegraphics[width=60mm,keepaspectratio]{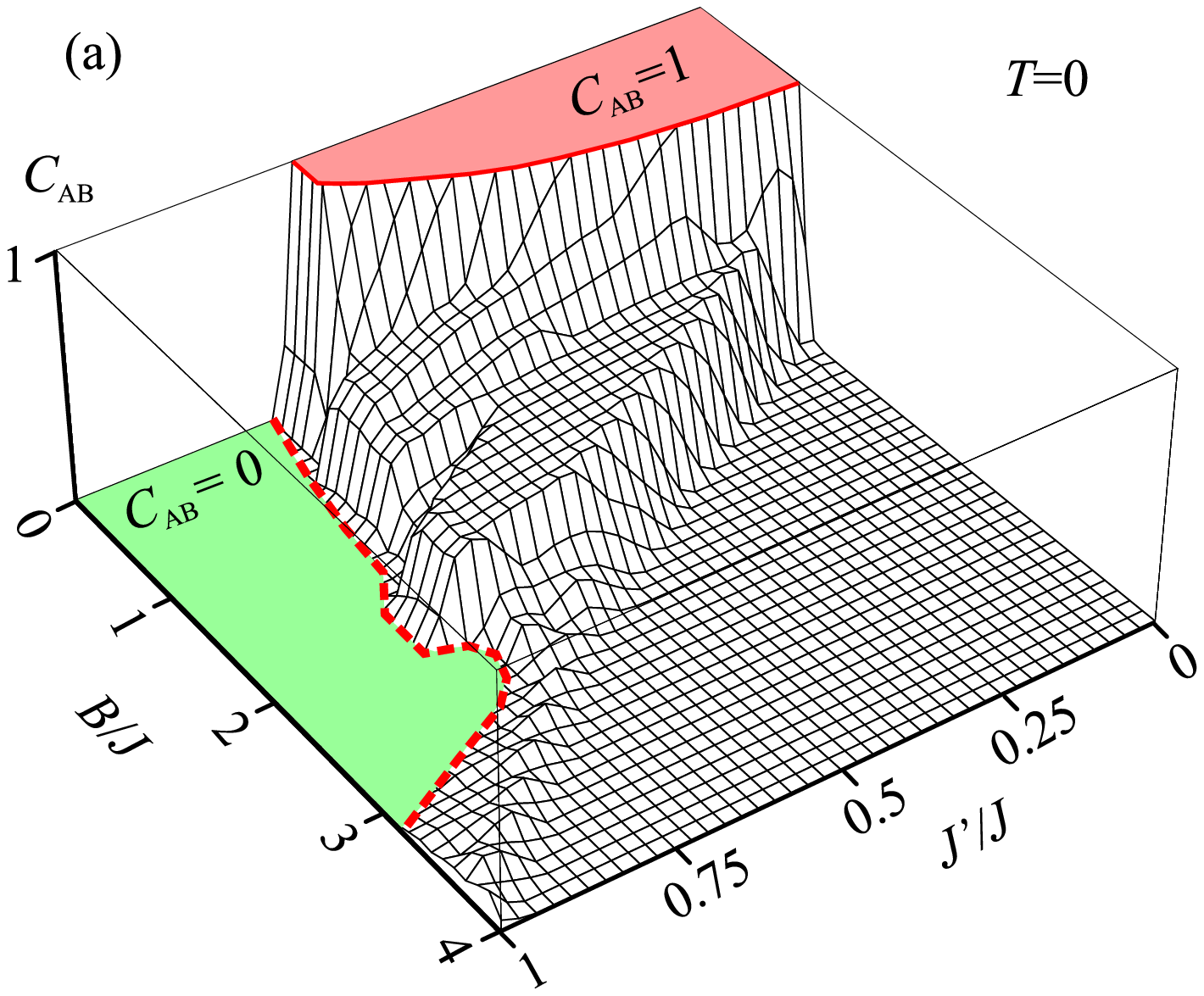}
\includegraphics[width=60mm,keepaspectratio]{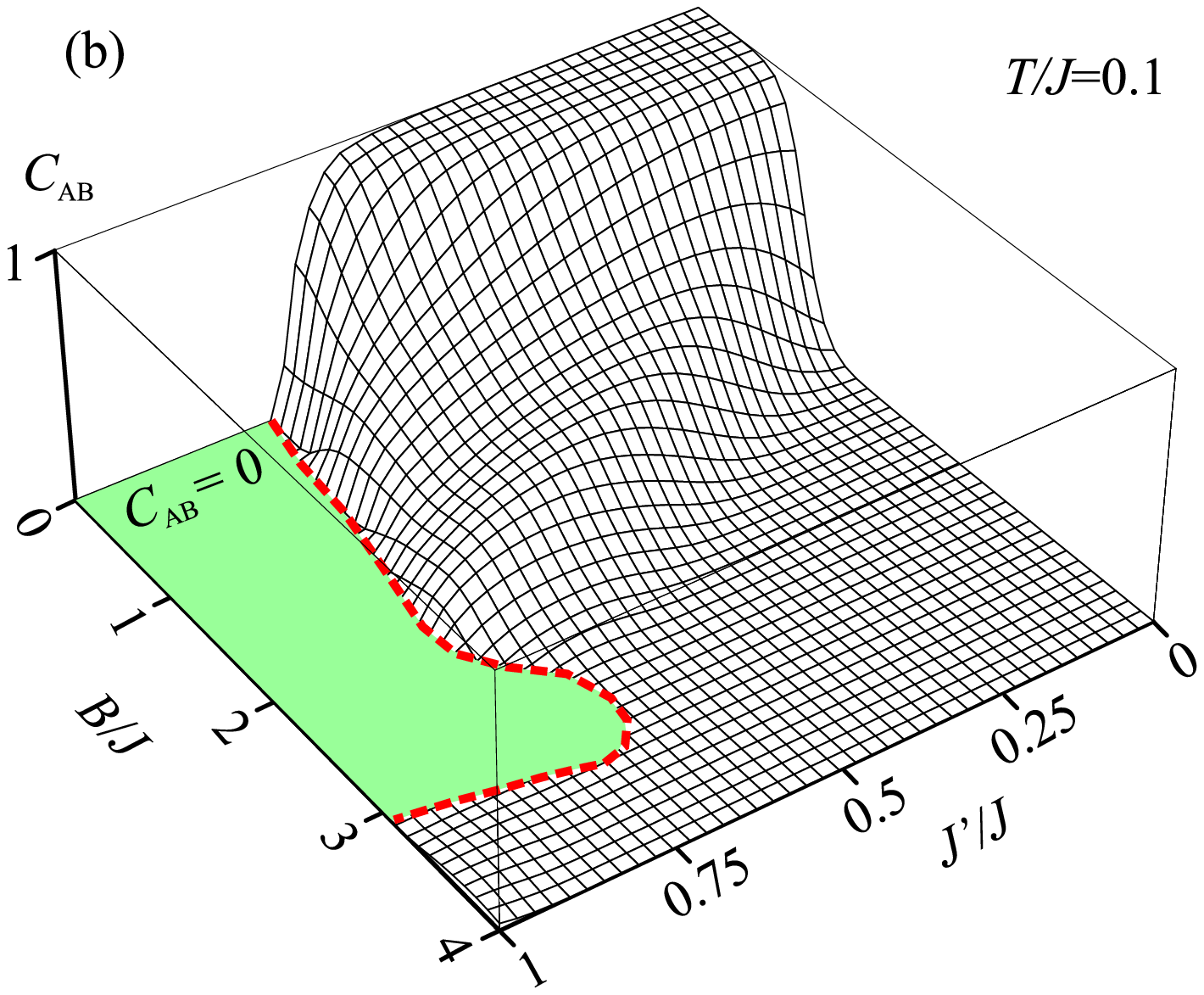}\\
\includegraphics[width=60mm,keepaspectratio]{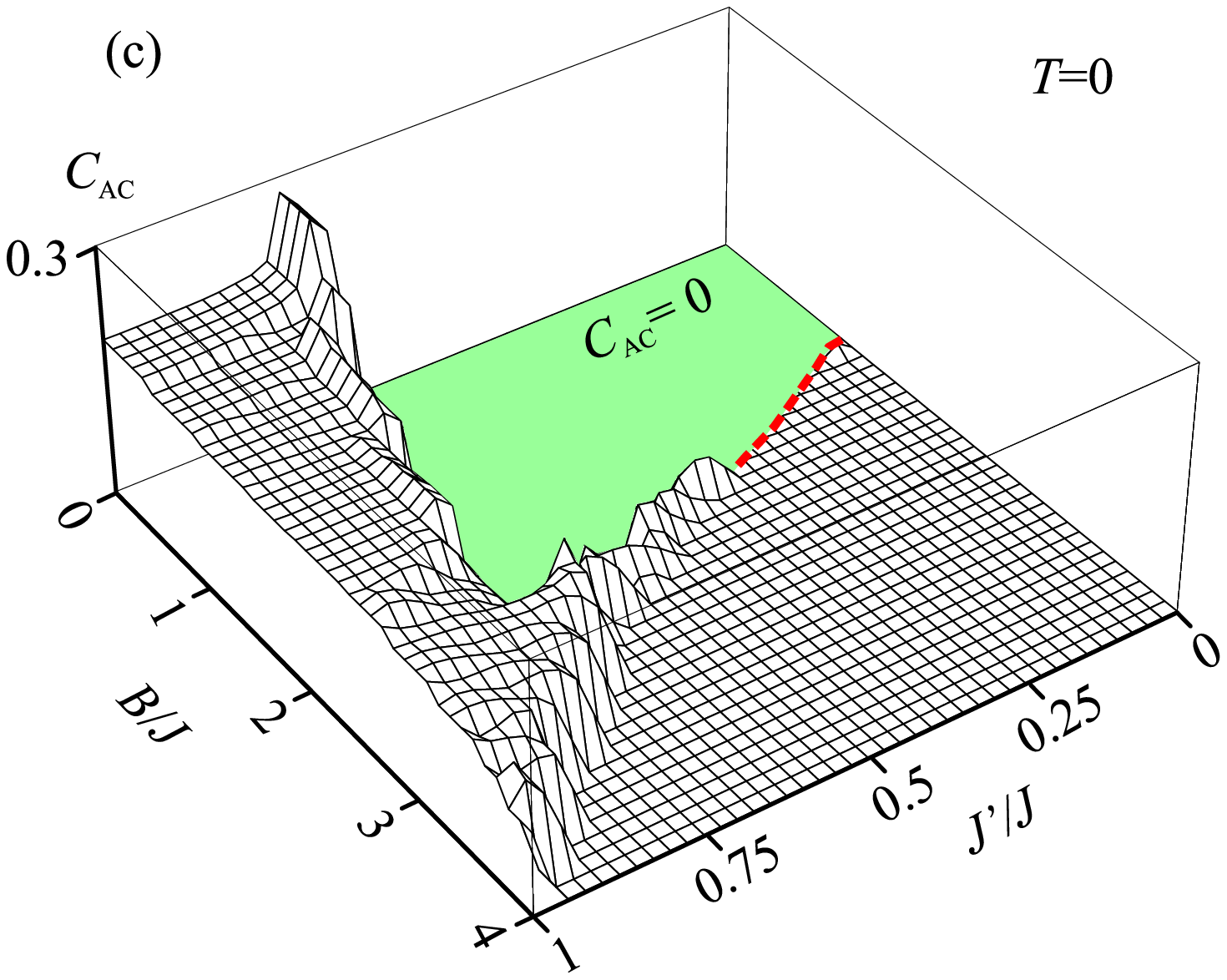}
\includegraphics[width=60mm,keepaspectratio]{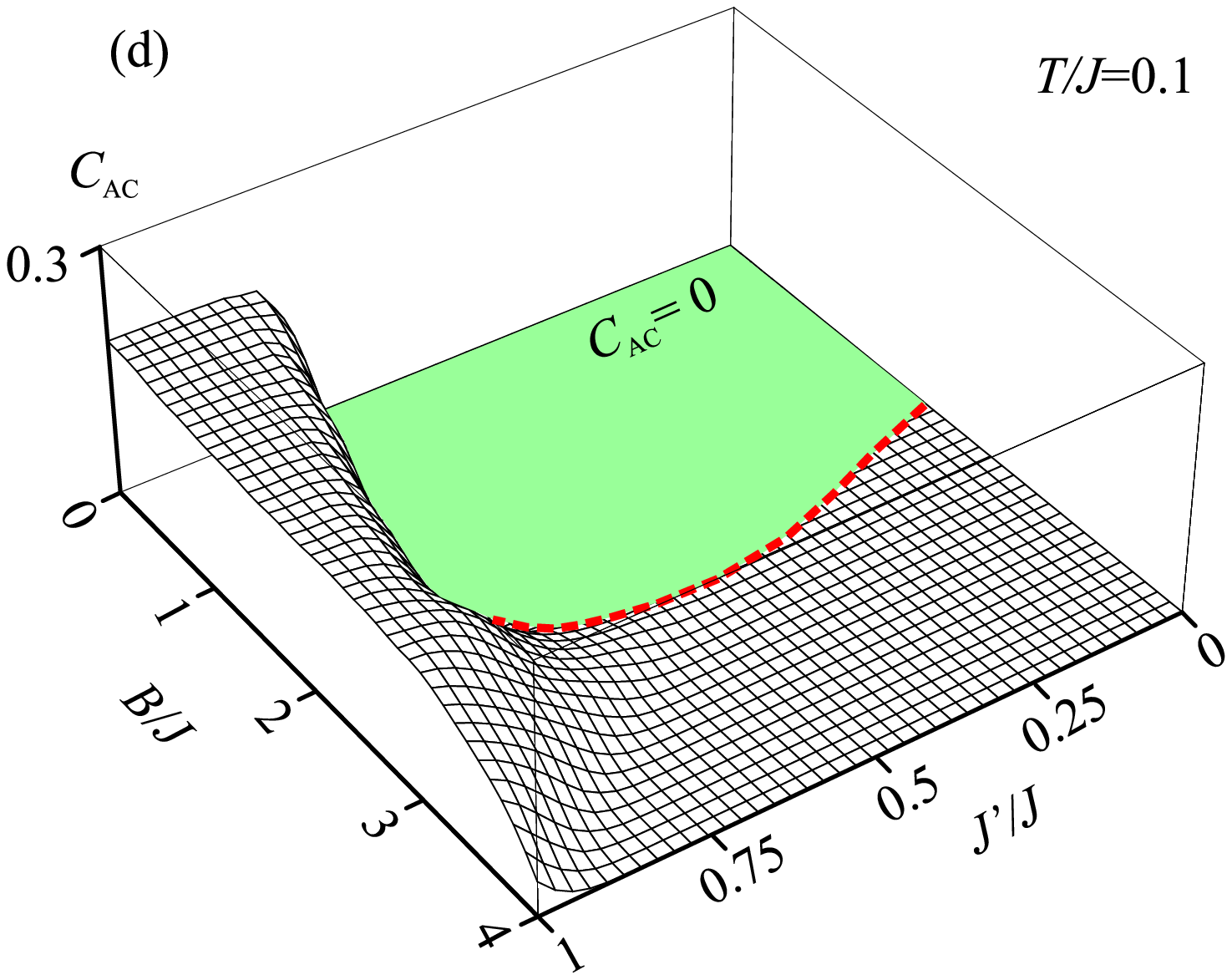}
\end{center}

\caption{(Color online) (a) Zero-temperature concurrence
$C_{\mathrm{A} \mathrm{B} }$ for a 20-site cluster for various $J'/J$
and $B/J$. Shaded area represents the regime of fully entangled
dimers, $C_{\mathrm{A} \mathrm{B} }=1$.  (b) The corresponding results
for $T/J=0.1$. (c) Next nearest concurrence $C_{\mathrm{A} \mathrm{C}
}$ for $T=0$, and (d) for $T/J=0.1$. Note qualitative and even
quantitative similarity with the tetramer results, Fig.~\ref{fig2}.
Dashed lines separate $C_{\mathrm{A}\mathrm{B}(\mathrm{C})}>0$ from
$C_{\mathrm{A}\mathrm{B}(\mathrm{C})}=0$.
}
\label{fig5}
\end{figure*}

\subsection{Numerical method}
We use the low-temperature Lanczos method \cite{ltlm} (LTLM), an
extension of the finite-temperature Lanczos method \cite{jj1} (FTLM)
for the calculation of static correlation functions at low
temperatures. Both methods are nonperturbative, based on the Lanczos
procedure of exact diagonalization and random sampling over different
initial wave functions. A main advantage of LTLM is that it accurately
connects zero- and finite-temperature regimes with rather small
numerical effort in comparison to FTLM. On the other hand, while FTLM
is limited in reaching arbitrary low temperatures on finite systems,
it proves to be computationally more efficient at higher
temperatures. A combination of both methods therefore provides
reliable results in a wide temperature regime with moderate
computational effort. We note that FTLM was in the past successfully
used in obtaining thermodynamic as well as dynamic properties of
different models with correlated electrons as are: the $t$-$J$
model,\cite{jj1} the Hubbard model,\cite{bonca02} as well as the SSM
model.\cite{jorge03,elshawish2}

In comparison with the conventional Quantum Monte Carlo (QMC) methods
LTLM possesses the following advantages: (i) it does not suffer from
the minus-sign problem that usually hampers QMC calculations of
many-electron as well as frustrated spin systems, (ii) the method
continuously connects the zero- and finite-temperature regimes, (iii)
it incorporates as well as takes the advantage of the symmetries of
the problem, and (iv) it yields results of dynamic properties in the
real time in contrast to QMC calculations where imaginary-time Green's
function is obtained. The LTLM (FTLM) is on the other hand limited to
small lattices which usually leads to sizable finite-size effects. To
account for these, we applied LTLM to different square lattices with
$N=8,16$ and 20 sites using periodic boundary conditions (we note that
next-larger system, $N=32$, was too large to be handled
numerically). Another drawback of the LTLM (FTLM) is the difficulty of
the Lanczos procedure to resolve degenerate eigenstates that emerge
also in the SSM.
%
%
In practice, this manifests itself in severe statistical fluctuations
of the calculated amplitude for $T\to 0$ since in this regime only a
few (degenerate) eigenstates contribute to thermal average. The
simplest way to overcome this is to take a larger number of random
samples $R\gg 1$, which, however, requires a longer CPU time. We have,
in this regard, also included a small portion of anisotropy in the SSM
(in the form of the anisotropic interdimer Dzyaloshinskii-Moriya interaction $D^z
\sum_{\{\rm{AC}\}} (S_A^xS_C^y-S_A^yS_C^x)$, $D^z/J\sim 0.01$),
which slightly splits the doubly degenerate single-triplet levels.
In this way, $R\sim 30$ per $S^z$ sector was enough for all
calculated curves to converge within $\sim 1\%$ for $T/J<1$. Here,
the number of Lanczos iterations $M=100$ was used along with the
full reorthogonalization of Lanczos vectors at each step.

\subsection{Entanglement}

Entanglement in the absence of magnetic field is most prominently
reflected in spin-spin correlation functions, {\it e.g.},
$\langle\textrm{\bf S}_{\mathrm{A}}\cdot \textrm{\bf
S}_{\mathrm{B}}\rangle $ and $\langle\textrm{\bf S}_{\mathrm{A}}\cdot
\textrm{\bf S}_{\mathrm{C}}\rangle $. In zero temperature limit due to
quantum phase transition at $J_c'$ these correlations change sign. In
Fig.~\ref{fig4} are presented renormalized spin-spin correlation
functions (for positive values identical to concurrence) as a function
of $J'/J$: (i) $C_{\mathrm{A} \mathrm{B} }>0$ in dimer phase and (ii)
$C_{\mathrm{A} \mathrm{C} }>0$ in the N{\' e}el phase. Critical $J'_c$
is indicated by asterisk. The results for $N=16$ are qualitatively and
quantitatively similar to the $N=20$ case presented here.  At finite
temperatures spin correlations are smeared out as shown in
Fig.~\ref{fig4} for various $T$. Limiting Heisenberg case, $J'\to
\infty$, is discussed in more detail in the next Section. $J'=0$ case
corresponds to the single dimer limit\cite{arnesen01} and Sec.
\ref{secII}.

Complete phase diagram of the SSM at $T=0$ but with finite magnetic
field can be classified in terms of concurrence instead of spin
correlations. In Fig.~{\ref{fig5}(a)$C_{\mathrm{A}
\mathrm{B}}$  is presented  as a function of $(J'/J,B/J)$ as in the case of a single
tetramer, Fig.~{\ref{fig2}(a). Presented results correspond to the
$N=20$ case, while $N=16$ system exhibits very similar structure (not
shown here). $N=8$ and $N=4$ cases are qualitatively similar, the main
difference being the value of critical $J'_c$ which increases with
$N$. Remarkable similarity between all these cases can be interpreted
by local physics in the regime of finite spin gap, $J'<J'_c$. Qubit
pairs are there completely entangled, $C_{\mathrm{A} \mathrm{B}}=1$,
and $C_{\mathrm{A} \mathrm{B}}\sim {1\over 2}$ for magnetic field
larger than the spin gap, but $B<J+2 J'$. For even larger $B$
concurrence approaches zero, similar to the $N=4$ case. Concurrence is
zero also for $J'>J'_c$, except along the $B\sim 4 J'$ line where weak
finite concurrence could be the finite size effect. Similar results
are found also for $N=16,8$ cases, and are most pronounced in the
$N=4$ case. At finite temperature the structure of concurrence is
smeared out [Fig.~\ref{fig5}(b)] similar to Fig.~\ref{fig2}(b).

Concurrence $C_{\mathrm{A} \mathrm{C}}$ corresponding to next-nearest
neighbors is, complementary to $C_{\mathrm{A} \mathrm{B}}$, increased
in the N\' eel phase of the diagram, Fig.~\ref{fig5}(c). The
similarity with $N=4$, Fig.~\ref{fig2}(c) is somewhat surprising
because in this regime long-range correlations corresponding to the
gapless spectrum of AFM-like physics are expected to change also short
range correlations. The only quantitative difference compared to $N=4$
is the maximum value of $C_{\mathrm{A} \mathrm{C}}\sim 0.3$ instead of
$0.5$ (beside the critical value $J'_c$ discussed in the previous
paragraph). Concurrence is very small for $B>J+2 J'$. At finite
temperatures fine fluctuations in the concurrence structure are
smeared out, Fig.~\ref{fig5}(d).

Temperature and magnetic field dependence of $C_{\mathrm{A}
\mathrm{B}}$ in the dimer phase is presented in Fig.~\ref{fig6}
for fixed $J'/J=0.4$.  Similarity with the corresponding $N=4$
tetramer case, Fig.~\ref{fig3}(a), is astonishing and is again the
consequence of local physics in the presence of a finite spin gap.
Finite size effects (in comparison with $N=16$ and $N=8$ cases)
are very small (not shown). Dashed line represents the borderline
of the $C_{\mathrm{A} \mathrm{B}}=0$ region: critical $T_c\approx
0.75 J$ valid for $B/J\lesssim 3$, that is in this regime nearly
independent of $B$, is slightly larger than in the single tetramer
case where its insensitivity to $B$ is even more pronounced.

\begin{figure}
\begin{center}
\includegraphics[width=60mm,keepaspectratio]{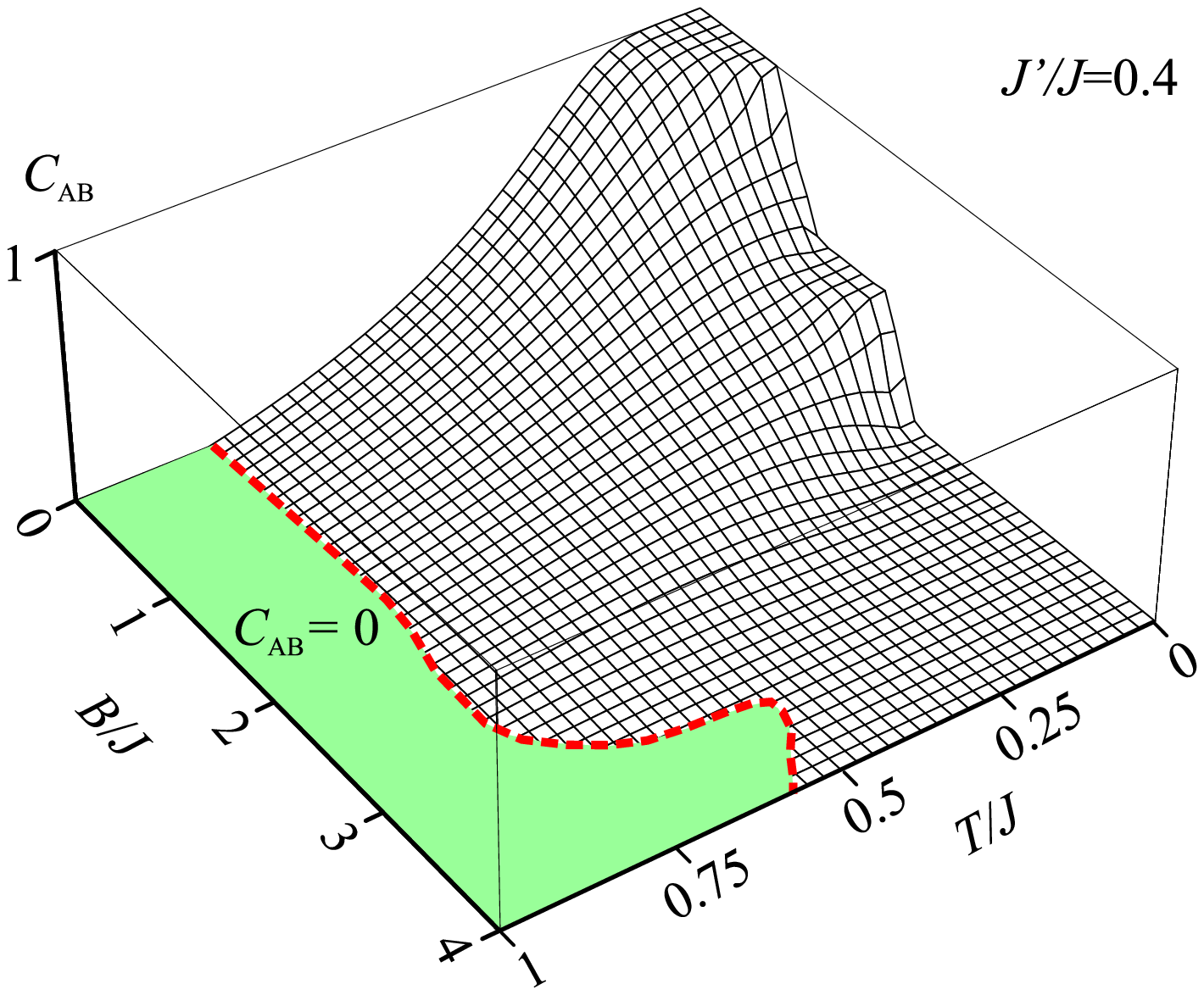}
\end{center}

\caption{(Color online) Temperature and magnetic field dependence of
$C_{\mathrm{A}\mathrm{B}}$ for $J'/J=0.4$ and $N=20$.  Note the
similarity with the corresponding tetramer results,
Fig.~\ref{fig3}(a). Dashed lines separate $C_{\mathrm{A}\mathrm{B}}>0$ from
$C_{\mathrm{A}\mathrm{B}}=0$.}

\label{fig6}
\end{figure}

\section{Heisenberg limit}
\label{secIV}

The concurrence corresponding to next-nearest neighbors in SSM,
$C_{\mathrm{A}\mathrm{C}}$, is non zero in the N\' eel phase for
$J'>J'_c$.  Typical result for concurrence in this regime (for
fixed $J'/J=1$) in terms  of temperature and magnetic field is
presented in Fig.~\ref{fig7}(a). At zero temperature the
concurrence is zero for $B>4J'$ [compare with Fig.~\ref{fig2}(c)
and Fig.~\ref{fig5}(c)].

In the limit $J=0$ the model simplifies to the AFM Heisenberg model on
a square lattice of $N$ sites,
\begin{equation}
  H_{\rm AC}=J' \sum_{\{\rm AC\}}{\bf S}_i\cdot {\bf S}_j-B\sum_{i=1}^{N} S_i^z.
\label{heisenberg}
\end{equation}
Several results for this model have already been presented for very
small clusters\cite{wang012,cao05,zhang05}, however the temperature
and magnetic field dependence of the concurrence for systems with
sufficiently large number of states and approaching thermodynamic
limit has not been presented so far.

\begin{figure}
\begin{center}
\includegraphics[width=60mm,keepaspectratio]{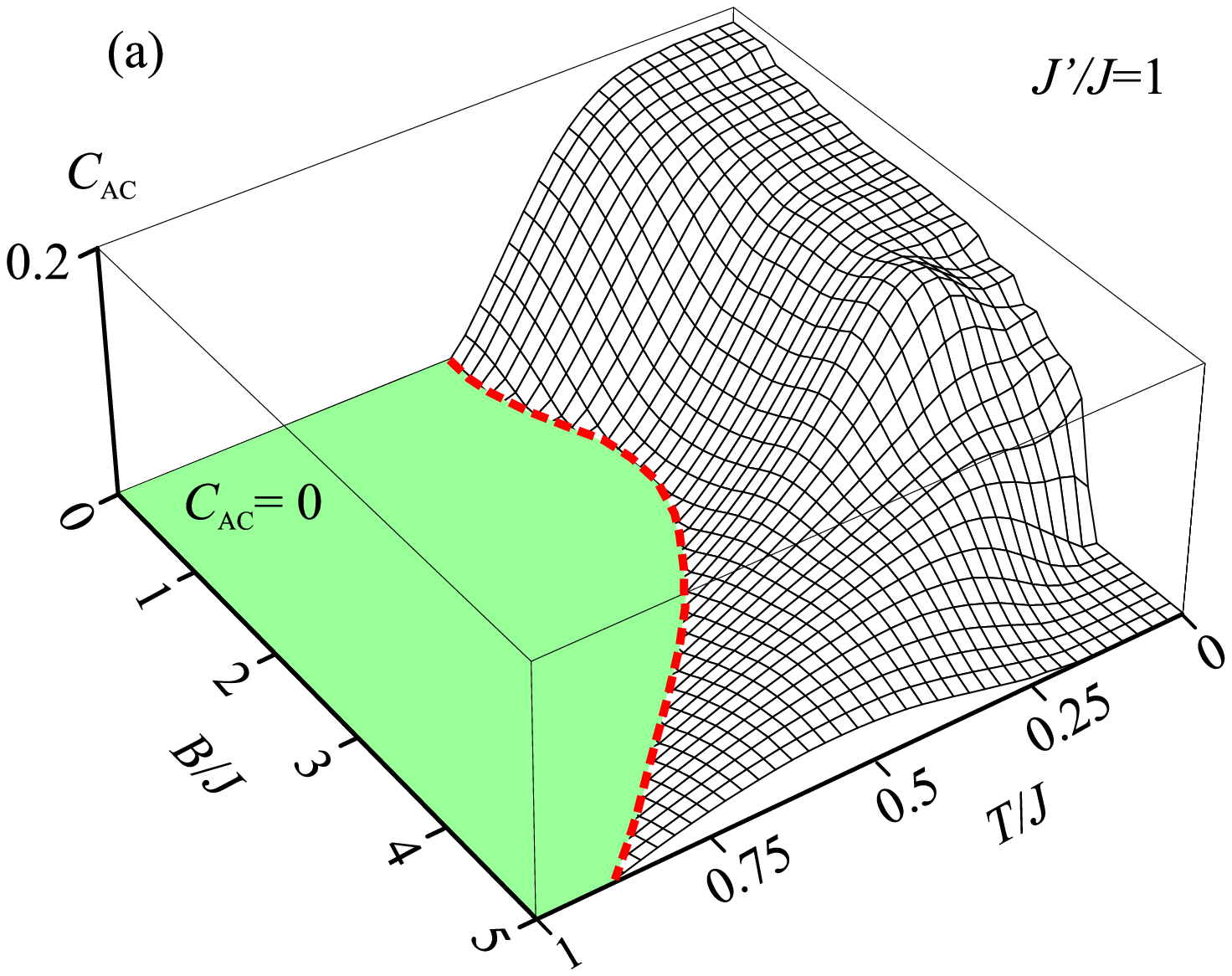}\\
\includegraphics[width=60mm,keepaspectratio]{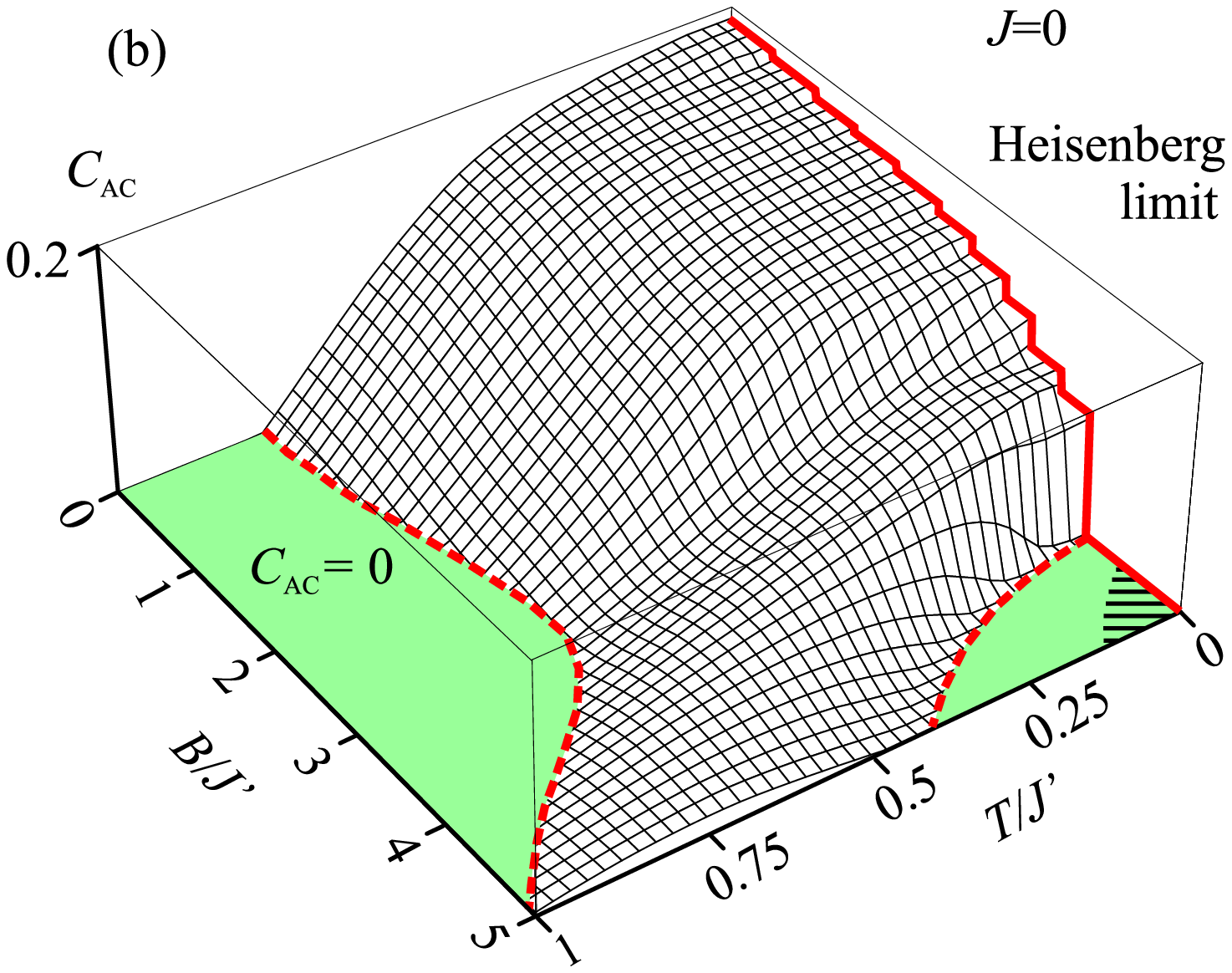}
\end{center}
\caption{(Color online) (a) Next nearest neighbor concurrence
$C_{\mathrm{A}\mathrm{C}}$ for $J'=J$. (b) Heisenberg lattice result
as a special case of the SSM, $J=0$.  Shaded region represents
$C_{\mathrm{A}\mathrm{C}}=0$. In the line shaded region (low
finite temperature and large magnetic field) our numerical results
set only the upper limit
$C_{\mathrm{A}\mathrm{C}}<5\cdot 10^{-4}$.  
Dashed lines separate $C_{\mathrm{A}\mathrm{C}}>0$ from
$C_{\mathrm{A}\mathrm{C}}=0$.}

\label{fig7}
\end{figure}

\begin{figure}
\begin{center}
\includegraphics[width=75mm,keepaspectratio]{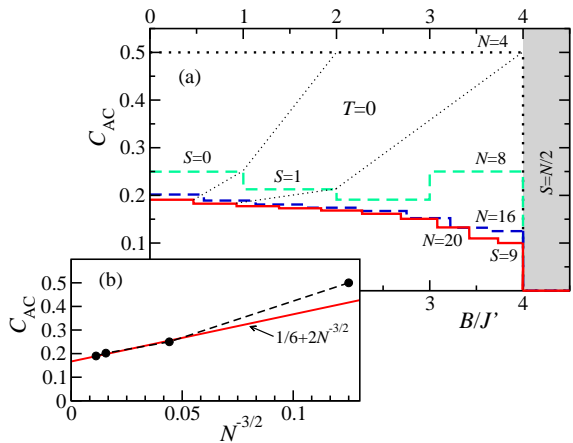}
\end{center}

\caption{(Color online) (a) Zero-temperature concurrence
$C_{\mathrm{A}\mathrm{C}}$ in the Heisenberg limit as a function of
$B/J'$ and for various $N=4$, $8$, $16$, $20$.  Sections with
different total spin values are additionally labeled.  (b) Finite-size
scaling of concurrence in the absence of magnetic field. Full line
represents the fit corresponding to Ref.~\onlinecite{manousakis},
$C_{\mathrm{A}\mathrm{C}}\approx {1\over 6}+2N^{-3/2}$.  }

\label{fig8}
\end{figure}

In Fig.~\ref{fig7}(b) we further  presented temperature and
magnetic field dependence of concurrence for the Heisenberg model
for $N=20$ (results for $N=16$ are quantitatively similar, but not
shown here). Temperature and magnetic field dependence of
$C_{\mathrm{A}\mathrm{C}}$ exhibits peculiar semi-island shape
where at fixed value of $B$ the concurrence increases with
increasing temperature. This effect is to some extent seen in all
cases and is the consequence of exciting local singlet states,
which do not appear in the ground state. At $T\to 0$ finite steps
with increasing $B$ correspond to gradual transition from the
singlet ground state to totally polarized state with total spin
$S=10$ and vanishing concurrence. This is in more detail presented
in Fig.~\ref{fig8}(a) for various $N=4, 8, 16, 20$. At $B=0$ and
for $N=20$ we get $C_{\mathrm{A}\mathrm{C}}=0.19$. It is
interesting to compare this results with the known finite-size
analysis scaling for the ground state energy of the Heisenberg
model \cite{manousakis}. The same scaling gives
$C_{\mathrm{A}\mathrm{C}}\approx {1 \over 6}+2N^{-3/2}$. Our
finite-size scaling, Fig.~\ref{fig8}(b), is in perfect agreement
with this result for $N\to \infty$ at $T=0$ and $B=0$.

In the opposite limit of high magnetic fields, the vanishing
concurrence $C_{\rm AC}=0$, is observed for $B$ above the critical
value $B_c=4J'$ for all system sizes shown in Fig.~\ref{fig8}(a).
This result can be deduced also analytically. Since in a fully
polarized state $C_{\rm AC}=0$, this $B_c$ actually denotes a
transition from $S_1=N/2-1$ to $S_0=N/2$ ferromagnetic ground
state with energy $E_0=N(J'-B)/2$. The energy of the one-magnon
excitation above the ferromagnetic ground state is given by the
spin wave theory, which is in this case exact, as
$E_1=E_0-J'(2-\cos k_x a-\cos k_y a)+B$, where $(k_x,k_y)$ is the
magnon wave vector and $a$ denotes the lattice spacing. Evidently,
a transition to a fully polarized state occurs precisely at
$B_c=4J'$ at $(\pi/a,\pi/a)$ point in the one-magnon Brillouin
zone.

\section{Summary}
\label{secV}

The aim of this paper was to analyze and understand how concurrence
(and related entanglement) of qubit pairs (dimers) is affected by
their mutual magnetic interactions. In particular, we were interested
in a planar array of qubit dimers described by the Shastry-Sutherland
model. This model is suitable due to very robust ground state composed
of entangled qubit pairs which breaks down by increasing the
interdimer coupling. It is interesting to study both, the entanglement
between nearest and between next-nearest spins (qubits) at finite
temperature and magnetic field. The results are based on numerical
calculations using low-temperature Lanczos methods on lattices of 4, 8,
16 and 20 sites with periodic boundary conditions.

A comprehensive analysis of concurrence for various parameters
revealed two general conclusions:

(1) For a weak coupling between qubit dimers, $J'<J'_c$, qubit pairs
are locally entangled in accordance with the local nature of the dimer
phase. This is due to a finite singlet-triplet gap (spin gap) in the
excitation spectrum that is a consequence of strong geometrical
frustration in magnetic couplings. The regime of fully entangled
neighbors perfectly coincides with the regime of finite spin gap as
presented in Fig.~\ref{fig9}. Calculated lines for various system
sizes $N$ in Fig.~\ref{fig9}(a) denote regions (shaded for $N=20$) in
the $(J'/J,B/J)$ plane where $C_{\mathrm{A}\mathrm{B}}=1$ at
$T=0$. In the lower panel [Fig.~\ref{fig9}(b)] the lines
represent the energy gap $E_1-E_{\mathrm{GS}}$ between the first
excited state with energy $E_1$ with total spin projection $S^z=1$ and the ground
state with energy $E_{\mathrm{GS}}$ and total
spin projection $S^z=0$, calculated for $B=0$. For $J'<J'_c$
(full lines) $E_1-E_{\mathrm{GS}}$ corresponds to the value of the
spin gap. With an increasing magnetic field the spin gap closes
(shaded region for $N=20$) and eventually vanishes at the
$C_{\mathrm{A}\mathrm{B}}=1$ border line. Shaded regions in
Figs.~\ref{fig9}(a),(b) therefore coincide. Note also that the results
for $N=16$ and 20 sites differ mainly in $J'_c$.

As a consequence of finite spin gap and local character of
correlations it is an interesting observation that even $N=4$ results
as a function of temperature and magnetic field qualitatively
correctly reproduce $N=20$ results in the regime of $J'<J'_c$. The
main quantitative difference is in a renormalized value of $J'_c=J/2$
for $N=4$, as is evident from the comparison of
Figs.\ref{fig2},\ref{fig3} and Figs.\ref{fig5},\ref{fig6}. This
similarity of the results appears very useful due to the fact that
concurrence for tetrahedron-like systems ($N=4$) is given analytically
(Sec. \ref{secII}).

(2) In the opposite, strong interdimer coupling regime, $J'>J'_c$, the
excitation spectrum is gapless and the concurrence between
next-nearest qubits, $C_{\mathrm{A}\mathrm{C}}$, exhibits a similar
behavior as in the antiferromagnetic Heisenberg model $J/J'\to 0$. Our
$B=0$ results coincide with the known result extrapolated to the
thermodynamic limit $C_{\mathrm{A}\mathrm{C}}\approx {1 \over 6}$. In
finite magnetic field and $T=0$ the concurrence vanishes at $B_c=4J'$
when the system becomes fully polarized (ground state with the total
spin $S=N/2$). However, at elevated temperatures the concurrence
increases due to excited singlet states and eventually drops to zero
at temperatures above $T_c\approx J'$.

We can conclude with the observation that our analysis of concurrence
and related entanglement between qubit pairs was also found to be a
very useful measure for classifying various phases of the
Shastry-Sutherland model. As our numerical method is based on
relatively small clusters, we were unable to unambiguously determine
possible intermediate phases of the model in the regime $J' \sim
J'_c$, but we believe that concurrence will prove to be a useful probe
for the classification of various phases also in this regime using
alternative approaches. However, we were able to sweep through all
other dominant regimes of the parameters including finite temperature
and magnetic field.

\section{Acknowledgments}

The authors acknowledge J. Mravlje for useful discussions and the
support from the Slovenian Research Agency under Contract No. P1-0044.

\begin{figure}
\begin{center}
\includegraphics[width=75mm,keepaspectratio]{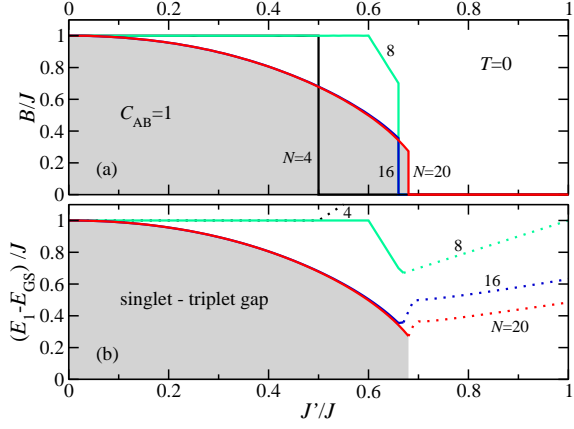}
\end{center}

\caption{(Color online) (a) Zero-temperature $C_{\mathrm{A}\mathrm{B}}=1$
region in the plane $(J'/J,B/J)$ for various $N$. (b) The
corresponding spin gap at $B=0$ (the energy of the lowest total
$S^z=1$ state relative to the ground state energy).}

\label{fig9}
\end{figure}

\appendix
\section{Eigenenergies and eigenvectors for periodically coupled
two qubit dimers}
\noindent
Consider two qubit dimers coupled into a tetramer and described with
the Hamiltonian Eq.~(\ref{habcd}) and Fig.~\ref{fig1}(a). The model is
exactly solvable in the separate $\{S,S^z\}$ subspaces corresponding
to different values of the total spin $S$ and its $z$ component
$S^z$. Following the abbreviations for singlet and triplet states on
nearest-neighbor (dimer) sites $i$ and $j$,
\begin{eqnarray}
   |s_{ij}\rangle &=& \frac{1}{\sqrt{2}}\, |\!\uparrow_i\downarrow_j -
   \downarrow_i\uparrow_j \rangle,\nonumber\\
   |t_{ij}^0\rangle &=& \frac{1}{\sqrt{2}}\, |\!\uparrow_i\downarrow_j +
   \downarrow_i\uparrow_j \rangle,\nonumber\\
   |t_{ij}^+\rangle &=& |\!\uparrow_i\uparrow_j \rangle,\nonumber\\
   |t_{ij}^-\rangle &=& |\!\downarrow_i\downarrow_j \rangle,
\end{eqnarray}
the resulting eigenstates $|\phi_k\rangle$ and eigenenergies $E_k$
corresponding to the hamiltonian Eq.~(\ref{habcd}) are:
\allowdisplaybreaks
\begin{align}
   S=0,&\ S^z=0:\nonumber\\
   |\phi_1\rangle &= |s_{\rm AB}\rangle |s_{\rm CD}\rangle,\nonumber\\
   E_1 &= -3J/2,\\[0.15cm]
   |\phi_2\rangle &= \frac{1}{\sqrt{3}}\big(
   -|t_{\rm AB}^0\rangle |t_{\rm CD}^0\rangle +
   |t_{\rm AB}^+\rangle |t_{\rm CD}^-\rangle +
   |t_{\rm AB}^-\rangle |t_{\rm CD}^+\rangle\big)\nonumber\\
   E_2 &= J/2-4J'.\\\nonumber\\
   S=1,&\ S^z=-1:\nonumber\\
   |\phi_3\rangle &= |s_{\rm AB}\rangle |t_{\rm CD}^-\rangle,\nonumber\\
   |\phi_4\rangle &= |t_{\rm AB}^-\rangle |s_{\rm CD}\rangle,\nonumber\\
   E_{3,4} &= -J/2-B,\\[0.15cm]
   |\phi_5\rangle &= \frac{1}{\sqrt{2}}\big(
   |t_{\rm AB}^0\rangle |t_{\rm CD}^-\rangle -
   |t_{\rm AB}^-\rangle |t_{\rm CD}^0\rangle\big),\nonumber\\
   E_5 &= J/2-2J'-B.\\\nonumber\\
   S=1,&\ S^z=0:\nonumber\\
   |\phi_6\rangle &=|s_{\rm AB}\rangle |t_{\rm CD}^0\rangle,\nonumber\\
   |\phi_7\rangle &=|t_{\rm AB}^0\rangle |s_{\rm CD}\rangle,\nonumber\\
   E_{6,7} &= -J/2,\\[0.15cm]
   |\phi_8\rangle &= \frac{1}{\sqrt{2}}\big(
   |t_{\rm AB}^+\rangle |t_{\rm CD}^-\rangle -
   |t_{\rm AB}^-\rangle |t_{\rm CD}^+\rangle\big),\nonumber\\
   E_8 &= J/2-2J'.\\\nonumber\\
   S=1,&\ S^z=1:\nonumber\\
   |\phi_9\rangle &= |s_{\rm AB}\rangle |t_{\rm CD}^+\rangle,\nonumber\\
   |\phi_{10}\rangle &= |t_{\rm AB}^+\rangle |s_{\rm CD}\rangle,\nonumber\\
   E_{9,10} &= -J/2+B,\\[0.15cm]
   |\phi_{11}\rangle &= \frac{1}{\sqrt{2}}\big(
   -|t_{\rm AB}^0\rangle |t_{\rm CD}^+\rangle +
   |t_{\rm AB}^+\rangle |t_{\rm CD}^0\rangle\big),\nonumber\\
   E_{11} &= J/2-2J'+B.\\\nonumber\\
   S=2,&\ S^z=-2:\nonumber\\
   |\phi_{12}\rangle &= |t_{\rm AB}^-\rangle |t_{\rm CD}^-\rangle,\nonumber\\
   E_{12} &= J/2+2J'-2B.\\\nonumber\\
   S=2,&\ S^z=-1:\nonumber\\
   |\phi_{13}\rangle &= \frac{1}{\sqrt{2}}\big(
   |t_{\rm AB}^0\rangle |t_{\rm CD}^-\rangle +
   |t_{\rm AB}^-\rangle |t_{\rm CD}^0\rangle\big),\nonumber\\
   E_{13} &= J/2+2J'-B.\\\nonumber\\
   S=2,&\ S^z=0:\nonumber\\
   |\phi_{14}\rangle &= \frac{1}{2}\big(
   2 |t_{\rm AB}^0\rangle |t_{\rm CD}^0\rangle +
   |t_{\rm AB}^+\rangle |t_{\rm CD}^-\rangle +
   |t_{\rm AB}^-\rangle |t_{\rm CD}^+\rangle\big),\nonumber\\
   E_{14} &= J/2+2J'.\\\nonumber\\
   S=2,&\ S^z=1:\nonumber\\
   |\phi_{15}\rangle &= \frac{1}{\sqrt{2}}\big(
   |t_{\rm AB}^0\rangle |t_{\rm CD}^+\rangle +
   |t_{\rm AB}^+\rangle |t_{\rm CD}^0\rangle\big),\nonumber\\
   E_{15} &= J/2+2J'+B.\\\nonumber\\
   S=2,&\ S^z=2:\nonumber\\
   |\phi_{16}\rangle &= |t_{\rm AB}^+\rangle |t_{\rm CD}^+\rangle,\nonumber\\
   E_{16} &= J/2+2J'+2B.
\end{align}

\end{document}